# Constant force amplifies pulse solitons via polarization coupling in mechanical metamaterials

Zehuan Tang,[1] Yuanwen Gao[1,*]

[1]*Department of Mechanics and Engineering Science, College of Civil Engineering and Mechanics, Lanzhou University, Lanzhou 730000, Gansu, China*

Unlike kink solitons with nonzero topological charge, pulse solitons cannot be driven by the experimentally most easily implemented constant force because of their zero topological charge. Here, we focus on a vector soliton with two polarization components: a pulse soliton component (rotation) and a kink soliton component (translation). We apply a constant force at the mass centers of the units, thereby driving the translational component with a constant force. Numerical simulations show that energy injected into the translational component can be transferred simultaneously to the rotational component through coupling, thereby achieving equal amplification of positive and negative pulses. Next, we change the application of this constant force to eccentric loading and find that the initially symmetric amplification of positive and negative solitons gradually becomes asymmetric over time. These phenomena are all explained by a theoretical model based on the adiabatic approximation, which gives a quantitative relation for the amplitude evolution, while the higher-order correction terms arising from the coupling effect also explain the origin of asymmetric transmission under eccentric driving. Our results demonstrate that vector solitons possess a unique capability for pulse soliton amplification by a constant force through polarization coupling and that this coupling can further modulate the symmetry between positive and negative pulses.

## I. INTRODUCTION

Solitons are a class of nonlinear excitations that can maintain their localized profiles during propagation. Pulse solitons and kink solitons are two typical forms [1–5]. For the field variable describing a soliton, the pulse soliton deviates from the background in the core region, but returns to the same background value at both ends and therefore has zero topological charge; the kink soliton connects different far-field states and has nonzero topological charge [6]. The two types of solitons differ not only in their waveforms but also in whether they leave a net change in the field variable after propagation. This difference directly affects how external driving injects energy into solitons.

For kink excitations, there are various ways to control them. These include relatively simple constant force driving [7–9] and release of stored energy through transitions between asymmetric steady states [10–12], as well as slightly more complex driving by periodic external forces [13,14]. These schemes can all control kink solitons with nonzero topological charge. For pulse excitations, when their profiles remain approximately unchanged, zero topological charge causes the driving effects of constant and periodic forces to be canceled out. On the other hand, their identical far-field states also make it impossible to use the energy difference between the configurations to supply them with energy continuously. This makes it almost impossible for the common driving methods mentioned above to effectively control pulse solitons. The driving methods for pulse solitons are often more complex than those for kink solitons, with examples including gain/loss coefficients [15] and time-dependent multiplicative factors [16].

In comparison, the constant force may still be the simplest form of driving. However, as described in the preceding paragraph, its driving effect is limited by the far-field structure of the soliton. To find a way to drive pulse solitons with a constant force, we focus on an elastic vector soliton formed by the coupling of a pulse soliton and a kink soliton. Such solitons have been shown to exhibit an amplitude gap and anomalous collision behavior, demonstrating the important influence of polarization coupling on their propagation [17,18]. In addition to these properties, this polarization coupling also has an effect that has not been fully utilized, namely that it makes the energy of a vector soliton a coupled whole rather than two separate parts [17]. This provides an approach worth trying for a constant force that provides no gain to the pulse. Is it possible to use the different far-field structures of the two components, so that a constant force can continuously inject energy into the pulse through the kink.

To this end, in a chain of cross units connected by flexible hinges, we apply constant-force driving in the soliton core region. First, we apply the force at the mass centers of the units so that it directly does work on the translational component. Numerical results show that the input energy can continuously amplify the rotational pulse component through polarization coupling. Since the driving force acts on the displacement component, it does not change the symmetry of the rotational component. Positive and negative pulses with the same initial amplitude undergo equal amplification. Furthermore, we move the application point of the force away from the mass center, thereby simultaneously producing a driving force in the translational component and a driving torque in the rotational component. At this point, the symmetry between positive and negative pulses is broken, and differences begin to appear in their amplification. Our study shows that polarization coupling can allocate the energy that a constant force injects into the displacement kink component to the rotational pulse component, and can also make the loading position a means of adjusting the symmetry of amplification between positive and negative pulses.

## II. MECHANICAL MODEL

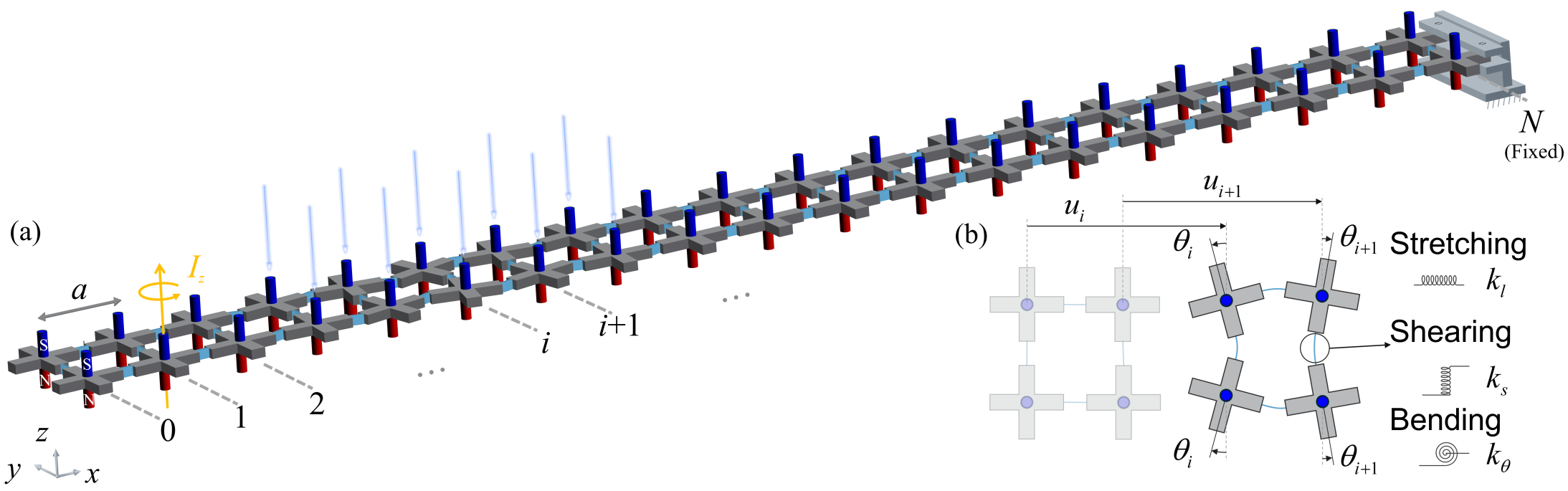

FIG. 1. (a) Schematic of the mechanical model, with the soliton core region covered by the magnetic field. (b) Schematic of the modeling approach. Each cross unit has two polarization components, a translational component $u_i$ and a rotational component $\theta_i$. The flexible hinges connecting the cross units are modeled as a combination of stretching, shearing, and torsional springs.

As shown in Fig. 1(a), we consider a chain assembled from cross units of mass $m$ with a spacing of $a$. The moment of inertia of each cross unit about the $z$ axis is $I_z$.

A permanent magnet rod passes through the mass center of each cross unit in the $z$ direction, with its S pole at the top and its N pole at the bottom. The magnetic field above the chain increases linearly in strength in the $x$ direction. It points in the negative $z$ direction and is given by $\boldsymbol{B} = \left(0{,}0, -(B_0 + Gx)\right)$, always covering $2H + 1$ nodes. The force exerted by this magnetic field on each permanent magnet rod is $\boldsymbol{f} = \nabla(\boldsymbol{m} \bullet \boldsymbol{B})$, where the magnetic moment $\boldsymbol{m} = (0{,}0, -m_z)$ is a vector pointing in the same direction as $\boldsymbol{B}$. Therefore, each magnet rod covered by the magnetic field experiences a rightward constant force $\boldsymbol{f} = (Gm_z, 0{,}0)$. The right end of the chain is fixed, and this constraint prevents the chain from translating as a whole.

As shown in Fig. 1(b), each cross unit has two polarization components, rotation and translation. This chain has been shown to support the propagation of a vector soliton [17]. In the chain subjected to the constant force described above, we also excite a vector soliton. At the same time, the magnetic field center follows the soliton center $X_0(T)$ while the field always covers $2H + 1$ nodes. The equations of motion of the chain are then written as

$$\ddot{U}_i = U_{i+1} - 2U_i + U_{i-1} - \frac{1}{2}\left(\cos\theta_{i+1} - \cos\theta_{i-1}\right) + F_i, \tag{1}$$

$$\begin{aligned} \frac{\ddot{\theta}_i}{\alpha^2} = &-K_\theta\left(\theta_{i+1} + 4\theta_i + \theta_{i-1}\right) + K_s \cos\theta_i \left(\sin\theta_{i-1} - 2\sin\theta_i + \sin\theta_{i+1}\right) \\ &- \sin\theta_i \left[4 - \cos\theta_{i-1} - 2\cos\theta_i - \cos\theta_{i+1} + 2\left(U_{i+1} - U_{i-1}\right)\right], \end{aligned} \tag{2}$$

where

$$F_i = \begin{cases} F, & i \in \left[\mathrm{Round}\left(X_0(T)\right) - H, \mathrm{Round}\left(X_0(T)\right) + H\right] \\ 0, & \text{else} \end{cases}. \tag{3}$$

Here, $\ddot{\blacksquare} = \frac{d^2}{dT^2}$. $U_i = \frac{u_i}{a}$ and $T = t\sqrt{\frac{k_l}{m}}$ are the dimensionless displacement and time, respectively. $K_s = \frac{k_s}{k_l}$ and $K_\theta = \frac{4k_\theta}{k_l a^2}$ are the dimensionless stiffness coefficients, and $\alpha = a\sqrt{\frac{m}{4I_z}}$ is a coefficient introduced by nondimensionalization. Their values are taken from Ref. [17], with $K_s = 0.02$, $K_\theta = 1.5 \times 10^{-4}$, and $\alpha = 1.8$. $F = \frac{f}{k_l a}$ is the dimensionless driving force. It is set to $F = 7.5 \times 10^{-6}$ in our numerical simulations and satisfies $F \ll 1$.

## III. NUMERICAL RESULTS

The dynamics of a chain with $N' = 400$ ($N' = N + 1$) units are numerically simulated under a driving force that follows the soliton. The excitation function applied at the 0-th node is the numerical solution of the perturbed continuum equations (see Appendix A for details). $H$ is chosen to be 13 so that the soliton core region always remains within the region driven by $F$. As shown by the numerical results in Fig. 2(a), the rightward constant force amplifies the amplitude of the positive pulse. When we change the form of excitation to excite a negative pulse with the same amplitude magnitude, the rightward constant force also amplifies the negative pulse soliton. The result is shown in Fig. 2(b). A comparison of Figs. 2(a) and 2(b) shows that the amplitude growth trajectories of the positive and negative pulses are identical. This shows that a constant force can amplify positive and negative pulses equally.

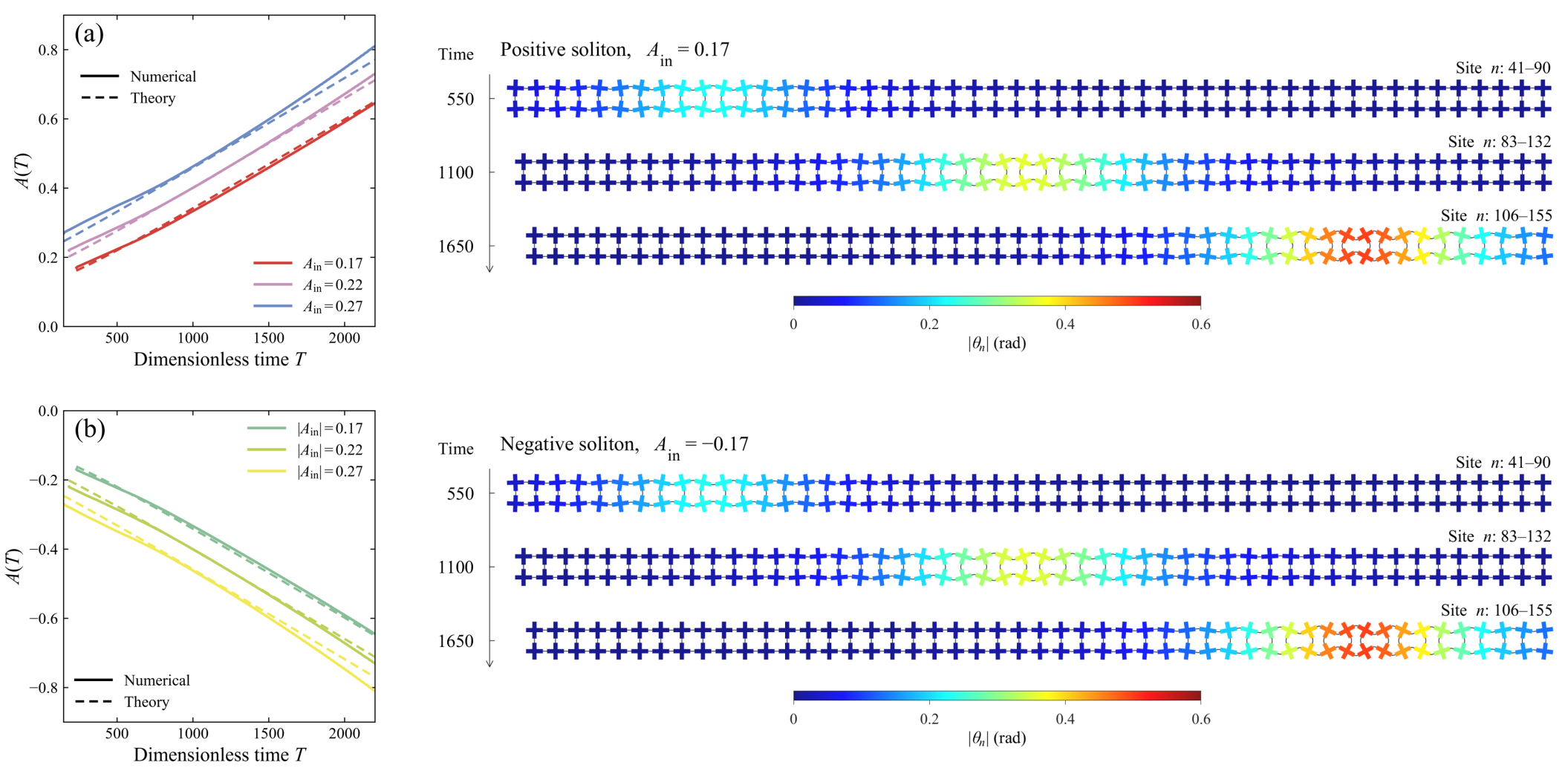

FIG. 2. (a) Plot of $A(T)$ for the positive pulse, with the actual propagation from numerical simulations shown on the right. (b) Plot of $A(T)$ for the negative pulse, with the actual propagation from numerical simulations shown on the right. The $A(T)$ curves in the two plots are perfectly symmetric.

The numerical results above show that a pulse soliton coupled to a kink soliton can possess some properties of the kink soliton. Specifically, it can be amplified by a constant force. However, it also differs from a kink soliton. When the sign of the solution changes, the direction of the driving (or gain) produced by the constant force does not change. By contrast, the driving direction of a kink soliton driven by a constant force depends on its topological charge. When the sign of the solution changes, the driving direction changes accordingly [19]. To better understand the numerical results above, we next develop a theory to clarify the physical mechanism underlying these results.

## IV. THEORETICAL MODEL

After Taylor expansion of Eqs. (1) and (2) and truncation of higher-order terms, the equations take the following forms in the continuum limit

$$U_{TT} = U_{XX} + \frac{1}{2}\left(\theta^2\right)_X + \begin{cases} F, & |X - X_0| \le H \\ 0, & |X - X_0| > H \end{cases}, \tag{4}$$

$$\frac{\theta_{TT}}{\alpha^2} = (K_s - K_\theta)\theta_{XX} - 6K_\theta\theta - 4U_X\theta - 2\theta^3. \tag{5}$$

Eq. (4) is multiplied by $2U_T$ and Eq. (5) by $\frac{\theta_T}{2}$ (these coefficients arise because Eq. (4) is obtained by dividing the original equation by 2, whereas Eq. (5) is obtained by multiplying the original equation by 2), and the two equations are then added and integrated to obtain the energy balance equation

$$\frac{dE}{dT} = 2\int_{-\infty}^{+\infty} F(X,T)U_T dX + \left[2U_T\left(U_X + \frac{1}{2}\theta^2\right) + \frac{K_s - K_\theta}{2}\theta_T\theta_X\right]\Big|_{-\infty}^{+\infty}, \tag{6}$$

where $E$ is the soliton energy and is given explicitly by

$$E(T) = \int_{-\infty}^{+\infty} U_T^2 + \frac{1}{4\alpha^2}\theta_T^2 + \frac{3K_\theta}{2}\theta^2 + \frac{1}{4}(K_s - K_\theta)(\theta_X)^2 + \left(U_X + \frac{1}{2}\theta^2\right)^2 dX. \tag{7}$$

The first term on the right-hand side of Eq. (6) is the power of the driving force. The second term on the right-hand side is a boundary term. Under the asymptotic boundary conditions $U_X|_{X\to\infty} = 0$, $\theta_X|_{X\to\infty} = 0$, and $\theta|_{X\to\infty} = 0$, the boundary term is zero. Therefore, the energy balance equation simplifies to

$$\frac{dE}{dT} = 2F\int_{X_0-H}^{X_0+H} U_T dX. \tag{8}$$

Since $F \ll 1$, it is assumed under the adiabatic approximation that the soliton driven by $F$ retains the forms of $U_{F=0}$ and $\theta_{F=0}$ [19]. The forms of $U_{F=0}$ and $\theta_{F=0}$ have been given in previous studies. The driven soliton solution can be written as

$$U = \frac{J(c)}{2}\left[1 - \tanh\frac{X - X_0(T)}{W(c)}\right], \tag{9}$$

$$\theta = \pm A(c)\,\mathrm{sech}\frac{X - X_0(T)}{W(c)}, \tag{10}$$

where $\dot{X}_0(T) = c$. Here, $c$ is no longer a constant but a function of $T$ given by $c = c(T)$. $J$, $A$, and $W$ are all functions of $c$, with $J(c) = \frac{6K_\theta W(c)}{c^2}$, $A(c) = \sqrt{\frac{6K_\theta(1-c^2)}{c^2}}$, and $W(c) = \sqrt{\frac{K_s - K_\theta - c^2/\alpha^2}{6K_\theta}}$. Substituting Eqs. (9) and (10) into Eq. (7) gives (see Appendix B for details)

$$E(T) = E_0(c) + cJJ'\dot{c} + \dot{c}^2\big(y(c)\big) + \dot{c}^2 J'^2 X_0, \tag{11}$$

where $\blacksquare' = \frac{d}{dc}$, and

$$E_0(c) = \frac{2K_\theta}{W(c)}\left[\frac{3(K_s - K_\theta)}{c^2} - (K_s - K_\theta) - \frac{2}{\alpha^2}\right]. \tag{12}$$

$E_0(c)$ can be regarded as the instantaneous energy of the soliton at speed $c$. Its energy density is nonzero only in the core region. Therefore, it depends only on the speed $c$ and is independent of the number of nodes actually occupied by the soliton. The following three terms are kinetic energy corrections arising from the time variation of $c$. The last term comes from the kink component. It shows that the kink component of the vector soliton differs from a kink soliton in a bistable structure. Its asymptotic value is not constrained by the potential wells. In the already transformed region (approximately $0 \sim X_0$ in a long chain), the continuously varying step height produces the correction term $\dot{c}^2 J'^2 X_0$, which grows with time. The change in the step in the already transformed region can be seen in Movie 1. The simulation animation in Movie 2 also shows the cross units in the already transformed region slowly translating to the right.

The expression for the power on the right-hand side of Eq. (8) is then obtained as

$$2F\int_{X_0-H}^{X_0+H} U_T dX = 2FJc\tanh\frac{H}{W} + 2FHJ'\dot{c}. \tag{13}$$

Substituting Eqs. (11) and (13) into Eq. (8) gives

$$\begin{aligned}\frac{dE_0(c)}{dc}\frac{dc}{dT} &+ (cJJ')'\left(\frac{dc}{dT}\right)^2 + cJJ'\frac{d^2c}{dT^2} + \frac{d\left[\dot{c}^2(y(c))\right]}{dT} + \dot{c}^2J'^2c + \frac{d\left[\dot{c}^2J'^2\right]}{dT}X_0 \\ &= 2FJc\tanh\frac{H}{W} + 2FHJ'\frac{dc}{dT}.\end{aligned} \tag{14}$$

Eq. (14) gives the differential equation for $c(T)$. To solve for $c(T)$, which is a slowly varying variable, the slow time $\tau = FT$ is introduced. The expansion of $c(T)$ on the slow time scale is

$$c(T) = c_0(\tau) + Fc_1(\tau) + O\left(F^2\right). \tag{15}$$

Here, $c(T)$ satisfies the initial condition $c(0) = c_{in}$. Since $c_{in}$ is independent of $F$, the initial values after the expansion in Eq. (15) are assigned as $c_0(0) = c_{in}$ and $c_1(0) = 0$. These initial values will be used to determine the solutions below. Substituting Eq. (15) into Eq. (14) gives

$$\begin{aligned}O\left(F^3\right) =& F\left[E_0{}'(c_0)\frac{dc_0}{d\tau} - 2J(c_0)c_0\tanh\frac{H}{W(c_0)}\right] \\ &+F^2\left[E_0{}''(c_0)c_1\frac{dc_0}{d\tau} + E_0{}'(c_0)\frac{dc_1}{d\tau} + c_0J(c_0)J'(c_0)\frac{d^2c_0}{d\tau^2} + \left(cJ(c)J'(c)\right)'\big|_{c=c_0}\left(\frac{dc_0}{d\tau}\right)^2\right. \\ &\left.+\left(\frac{dc_0}{d\tau}\right)^2\left[J'(c_0)\right]^2c_0 - 2\frac{d\left[J(c)c\tanh\frac{H}{W(c)}\right]}{dc}\Bigg|_{c=c_0}c_1 - 2HJ'(c_0)\frac{dc_0}{d\tau}\right].\end{aligned} \tag{16}$$

Eq. (16) separates the equations that need to be solved according to the powers of $F$. The equation at each order is obtained by setting the corresponding term in square brackets to zero. Since the time evolution in Fig. 2 is plotted using the actual time $T$, the first-order equation is rewritten in the original variable $T$ as

$$E_0{}'(c_0)\frac{dc_0}{dT} = 2FJ(c_0)c_0\tanh\frac{H}{W(c_0)}. \tag{17}$$

The second-order equation is

$$
\begin{aligned}
FE_0''(c_0)c_1\frac{dc_0}{dT} &+ FE_0'(c_0)\frac{dc_1}{dT} + c_0 J(c_0)J'(c_0)\frac{d^2c_0}{dT^2} + \left(cJ(c)J'(c)\right)'|_{c=c_0}\left(\frac{dc_0}{dT}\right)^2 \\
&+ \left(\frac{dc_0}{dT}\right)^2\left[J'(c_0)\right]^2 c_0 = 2F^2 \left.\frac{d\left[J(c)\text{ctanh}\frac{H}{W(c)}\right]}{dc}\right|_{c=c_0} c_1 + 2FHJ'(c_0)\frac{dc_0}{dT}.
\end{aligned}
\tag{18}
$$

Solving Eqs. (17) and (18) with the prescribed initial values gives $c_0(T)$ and $c_1(T)$. They are then combined according to $c(T) = c_0(T) + Fc_1(T)$ to obtain the final velocity $c(T)$. After the solution for $c(T)$ is obtained, the $A(c)$ relation is used to obtain the $A(T)$ curve. The theoretical $A(T)$ curve is plotted as the theoretical result in Fig. 2. It shows good agreement with the numerical simulations. This demonstrates the correctness of our theoretical model.

The theoretical model reveals the physical mechanism by which a constant force amplifies a pulse soliton. From the leading-order Eq. (17), it can be seen that a constant force does not amplify the pulse soliton by doing work on it directly. It relies on the fact that the energies of the pulse and the kink form a coupled whole. By doing work on the kink component, the constant force increases this coupled total energy and thereby increases the amplitude of the pulse. This energy input is not completely converted into the energy of the pulse, however. If $U_T$ in Eq. (8) is replaced with $\theta_T$, the constant drive acts directly on the pulse soliton. Since the first term of the integrand $\theta_T$ is an odd function,

$$
\theta_T = \pm\left\{\frac{Ac}{W}\operatorname{sech} z \tanh z + \dot{c}\left[A' \operatorname{sech} z + \frac{AW'}{W} z \operatorname{sech} z \tanh z\right]\right\}, \tag{19}
$$

the right-hand side of Eq. (17) becomes zero ($z = \frac{X - X_0(T)}{W(c)}$). The first-order equation then gives $\frac{dc_0}{dT} = 0$. $\frac{dc_0}{dT} = 0$ further causes the second-order equation to yield $\frac{dc_1}{dT} = 0$. This is because the source terms for $c_1$ are zero (these source terms include the second-order power term and other terms containing neither $c_1$ nor derivatives of $c_1$, and they all contain $\frac{dc_0}{dT}$ or its time derivative as a factor). The result is that a driving force is applied but produces no driving effect. The inability of a constant force to drive a pulse soliton is also explained by this result. In Appendix C, numerical results for a pulse soliton driven by a constant torque are presented. The nearly unchanged $A(T)$ confirms our theoretical analysis.

Furthermore, it can be seen that the governing equations (4) and (5) have the symmetry $\theta \to -\theta$, $U \to U$. The presence of $F$ does not break this symmetry. As a result, even when $\theta$ changes sign as $\theta \to -\theta$, $U$ remains unchanged. That is, the power integral $2F\int_{X_0-H}^{X_0+H} U_T dX$ that actually contributes is the same for positive and negative pulses. This not only explains the observation in Fig. 2 that the driving direction of the constant force is independent of the sign of the pulse, but also explains the completely symmetric $A(T)$ curves for positive and negative pulses.

At this point, we have systematically explained the physical mechanism by which a constant force drives pulse solitons. However, the current way of applying the driving force cannot break the symmetry inherent in Eqs. (4) and (5). This prevents us from going beyond the symmetry between positive and negative pulses. In the following section, we will offset the application point of the driving force on the cross units to break this symmetry, thereby achieving asymmetric amplification of positive and negative pulses.

## V. SYMMETRY BREAKING

### A. Asymmetric amplification

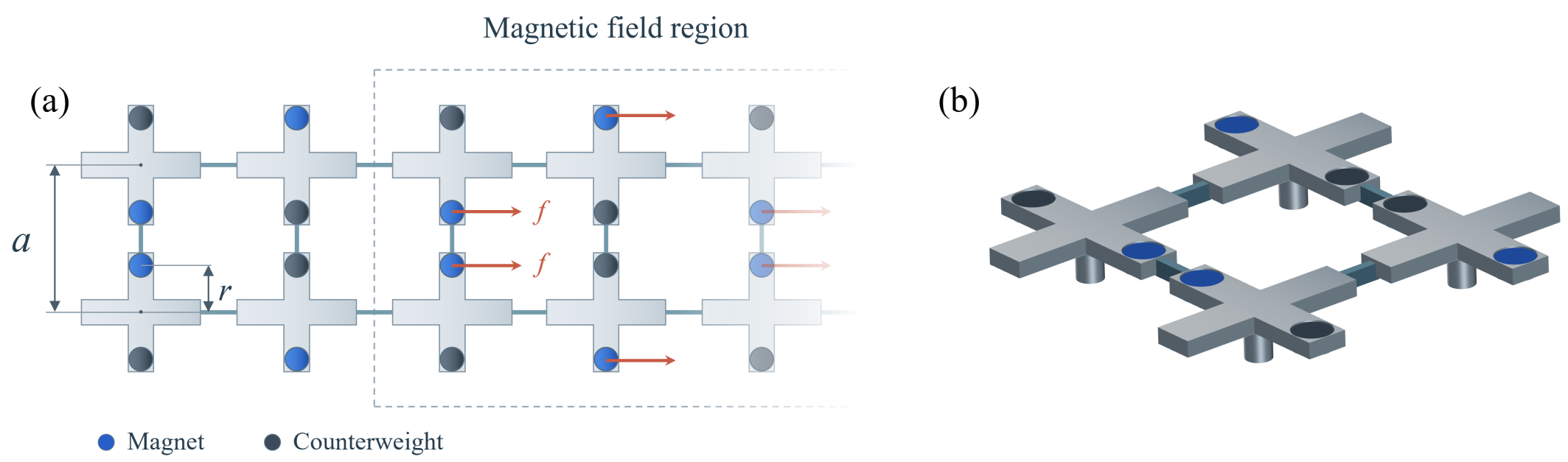

FIG. 3. (a) Schematic of the structure with eccentrically positioned magnets. The blue circles represent magnets. They are arranged alternately at the upper and lower ends of successive cross units. The black circles represent counterweights with the same mass as the magnets. These counterweights keep the mass centers of the cross units at their geometric centers. $r$ is the arm length of the cross unit minus the magnet radius. (b) Three-dimensional view of the structural units.

As shown in Fig. 3, the difference from the previous scheme in Fig. 1 is that the force application point is shifted to the end of the cross unit. Therefore, the cross units in the core region are now subjected to a combination of a rightward constant force and a torque. The governing equations of the system become

$$\ddot{U}_i = U_{i+1} - 2U_i + U_{i-1} - \frac{1}{2}\left(\cos\theta_{i+1} - \cos\theta_{i-1}\right) + F_i, \tag{20}$$

$$\begin{aligned}\frac{\ddot{\theta}_i}{\alpha^2} &= -K_\theta\left(\theta_{i+1} + 4\theta_i + \theta_{i-1}\right) + K_s \cos\theta_i \left(\sin\theta_{i-1} - 2\sin\theta_i + \sin\theta_{i+1}\right) \\ &\quad - \sin\theta_i\left[4 - \cos\theta_{i-1} - 2\cos\theta_i - \cos\theta_{i+1} + 2\left(U_{i+1} - U_{i-1}\right)\right] + 4bF_i\cos\theta_i\,.\end{aligned} \tag{21}$$

$b = \frac{r}{a}$, and $b = 0.4$ in our numerical simulations. The governing equations show that the $4bF_i\cos\theta_i$ term breaks the original $\theta \to -\theta,\ U \to U$ symmetry of the equations. When this term is active, the propagation of positive and negative pulses must be asymmetric. The numerical simulation results in Fig. 4(a) confirm that this term takes effect. For the same initial amplitude, with the amplitude evolution at $b = 0$ as the reference baseline, the positive pulse amplitude is always above the $b = 0$ baseline, whereas the negative pulse amplitude is always below it. Fig. 4(b) plots the amplitude differences of the positive and negative pulses using the $b = 0$ $A(T)$ curve as the baseline. It shows more clearly that the positive pulse propagates with a higher amplitude than the negative pulse.

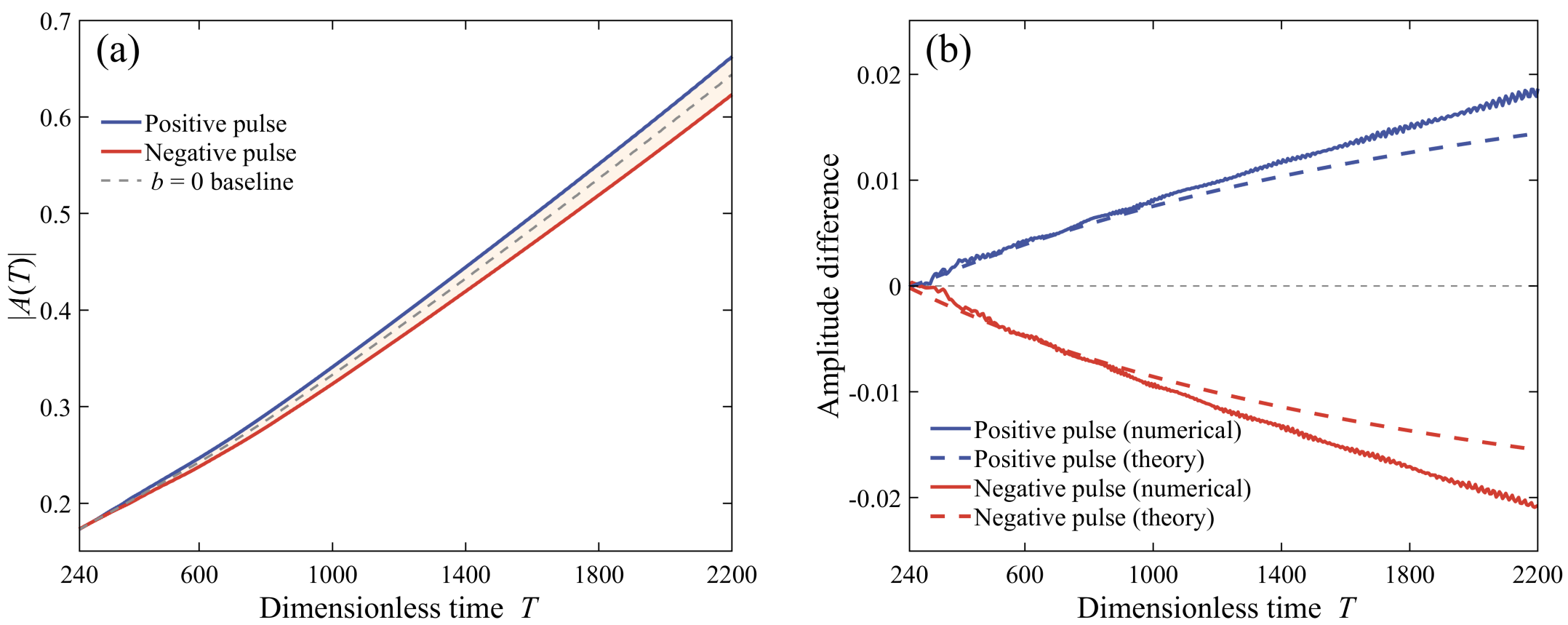

FIG. 4. (a) $|A(T)|$ curves for positive and negative pulses. The blue curve represents the positive pulse, the red curve represents the negative pulse, and the gray dashed curve is the $b = 0$ baseline. (b) Amplitude difference curves for the positive and negative pulses relative to the $b = 0$ baseline.

### B. Soliton stability

This asymmetric transmission can be explained by the higher-order theory developed in Section IV. Before turning to this theory, this section first uses the stability region of undriven solitons to determine the parameter range to be discussed in the subsequent theoretical model. As a broad restriction, we consider only the range with a nonzero lower amplitude bound, $0 < K_s - K_\theta < \frac{1}{\alpha^2}$ [17]. The material parameters in most existing studies also lie in this range [18,20–24]. Within this range, the boundary for a real-valued width ($W(c) = 0$) gives the point of criticality for the existence of the continuum soliton solution. Previous work used this lower bound to estimate the point of criticality for soliton stability in weakly discrete systems (here, we regard the discreteness of the actual system as a perturbation added to the continuum model, so the ability to propagate over long distances in a discrete system is interpreted as soliton stability). However, we performed numerical simulations of the following continuum equations

$$U_{TT} = U_{XX} + \frac{1}{2}\left(\theta^2\right)_X, \tag{22}$$

$$\frac{\theta_{TT}}{\alpha^2} = (K_s - K_\theta)\theta_{XX} - 6K_\theta\theta - 4U_X\theta - 2\theta^3. \tag{23}$$

The results show that $W(c) = 0$ is not the exact point of criticality for soliton stability. Specifically, a small perturbation was initially applied to the rotational component, $\theta = A_0 \mathrm{sech}\frac{X}{W_0}\left[1 - \varepsilon \mathrm{sech}^2\frac{X}{W_0}\right]$. $-A_0\varepsilon \mathrm{sech}^3\frac{X}{W_0}$ is the small perturbation introduced to test soliton stability, with $\varepsilon = 0.001$. The perturbation tends to zero far from the core region, which prevents it from producing uncontrolled effects as the chain length increases. With $A_0 = 0.13$ and $\alpha = 1.8$ fixed, soliton propagation was tested for different $K_s$ and $K_\theta$ values. The ratio of the amplitude at the right-end test point to the input reference amplitude, $\frac{A_{\mathrm{test}}}{A_0}$, was used to characterize soliton stability. The stability phase diagram is shown in Fig. 5(a), and propagation plots at representative points in the stable and unstable regions are shown in Fig. 5(b). Fig. 5(a) shows that criticality for soliton instability occurs below the black $W = 0$ curve. In other words, distortion in soliton propagation over long distances does not require reaching the boundary at which the real-valued width solution becomes invalid. It is certain that the numerically observed instability before $W = 0$ is not caused by discreteness [25]. Because we consider continuum equations, this early instability is a property of the continuum equations themselves.

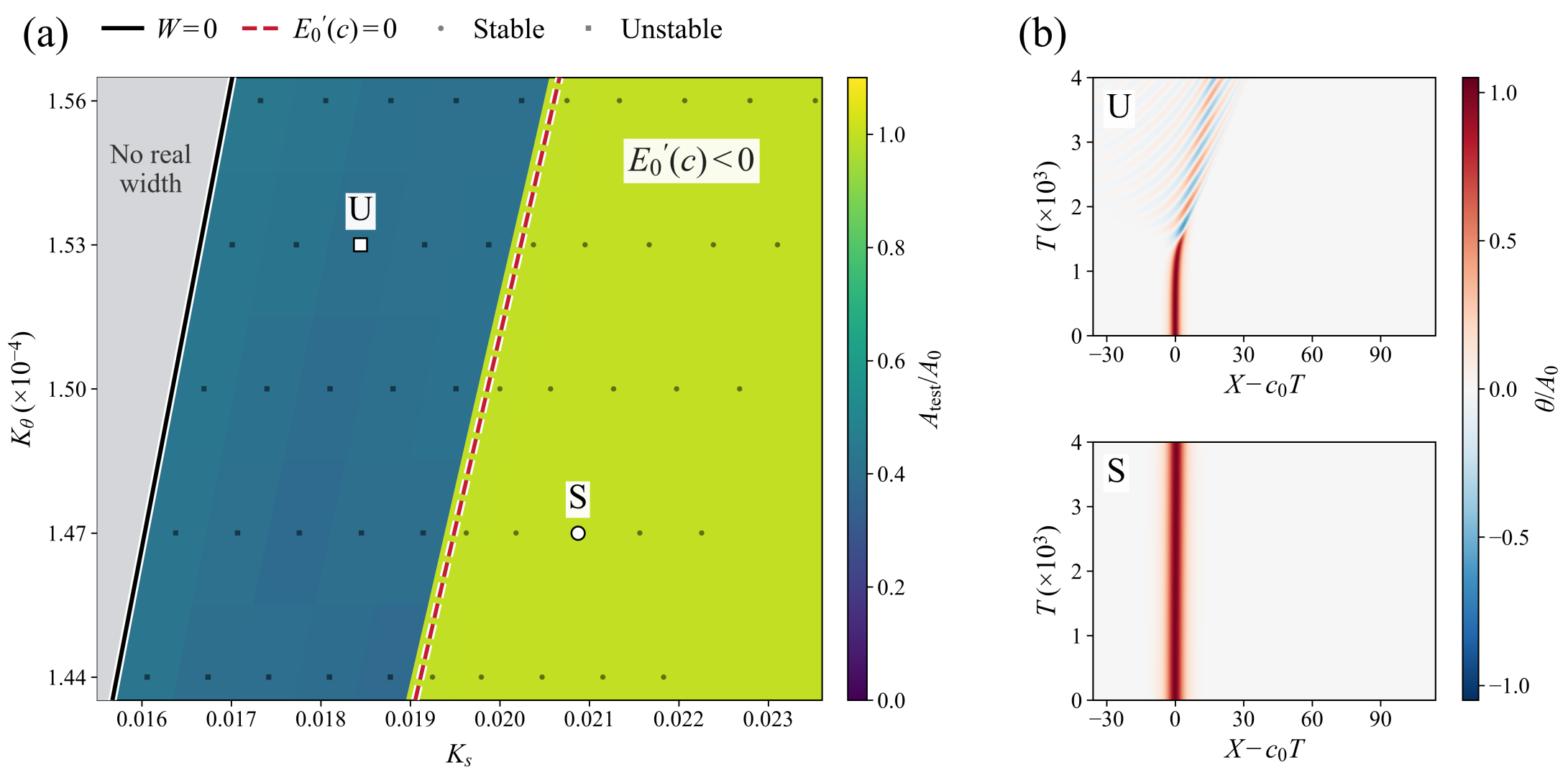

FIG. 5. (a) Soliton stability phase diagram for different parameter values. The fluorescent green region is the stable region for solitons, and the dark green region is the unstable region. The black solid line is the boundary given by the condition for a real-valued width $W = 0$, and the red dashed line is the boundary given by the stability condition $E_0'(c) = 0$. (b) Spatiotemporal plots of soliton propagation at two representative points in the stable and unstable regions. The horizontal coordinate $X - c_0 T$ is the soliton-centered reference coordinate based on $c_0$.

To explain this early instability, a stricter constraint than the condition for a real-valued width needs to be found. In comoving coordinates, the Hamiltonian of the system is

$$H_m(c) = E_0(c) - cp(c). \tag{24}$$

$p(c)$ is the momentum associated with the generalized coordinate $X_0(T)$ of the soliton center,

$$p(c) = \int_{-\infty}^{+\infty} \frac{\partial \mathscr{L}}{\partial \dot{X}_0(T)} dX = -\int_{-\infty}^{+\infty} \left( 2U_T U_X + \frac{\theta_T \theta_X}{2\alpha^2} \right) dX. \tag{25}$$

$\mathscr{L}$ is the Lagrangian density. $p(c)$ and $E_0(c)$ have the following relation

$$E_0{}'(c) = cp'(c). \tag{26}$$

Therefore,

$$H_m{}''(c) = -\frac{E_0{}'(c)}{c}. \tag{27}$$

For a locally well-posed system with conserved energy and momentum, if the spectrum of the operator corresponding to the quadratic form in the expansion of $H_m(c)$ contains one negative eigenvalue and one zero eigenvalue and the rest of the spectrum is positive and bounded away from zero (see Appendix D for details), then the condition for orbital stability of the system is given by $H_m{}''(c) > 0$ [26,27]. Therefore, for this continuum soliton system, the stable region should be given by $E_0{}'(c) < 0$, and the boundary is $E_0{}'(c) = 0$. $E_0{}'(c) = 0$ is plotted as the red dashed line in Fig. 5. It can be seen that this line almost gives the stability boundary of vector solitons.

In this section, we have given a stability range narrower than that given by the condition for a real-valued width. As indicated by this condition, our discussion will be within the parameter range $E_0{}'(c) < 0$. This is because this soliton solution will ultimately be applied to a (weakly) discrete system, and our system also has a constant-force perturbation. This requires the discussion to be carried out within the soliton stability range.

### C. Origin of asymmetry

For this case of eccentric loading, the theoretical model is obtained by adding the eccentric torque term to Eqs. (4) and (5)

$$U_{TT} = U_{XX} + \frac{1}{2}\left(\theta^2\right)_X + \begin{cases} F, & |X - X_0| \le H \\ 0, & |X - X_0| > H \end{cases}, \tag{28}$$

$$\frac{\theta_{TT}}{\alpha^2} = (K_s - K_\theta)\theta_{XX} - 6K_\theta\theta - 4U_X\theta - 2\theta^3 + \begin{cases} 4bF\cos\theta, & |X - X_0| \le H \\ 0, & |X - X_0| > H \end{cases}. \tag{29}$$

The energy balance equation is

$$\frac{dE}{dT} = 2F\int_{X_0-H}^{X_0+H} U_T dX + 2bF\int_{X_0-H}^{X_0+H} \theta_T \cos\theta\, dX. \tag{30}$$

The second term is the additional torque power term due to eccentric loading, and we focus on this term. The leading term of $\theta_T$ is an odd function and integrates to zero over a symmetric interval, which shows that the positive and negative pulses are completely identical in the first-order theory; namely, the result reduces to that for $b = 0$ in Section IV. However, unlike the case of pure torque, due to the presence of the constant force at the mass center, $\frac{dc_0}{dT}$ is nonzero. This gives rise to a second-order correction in this eccentrically driven system, whereas the second-order correction remains zero in the case of pure torque. The response of this eccentrically driven system can be regarded as a higher-order correction to the $b = 0$ system, and the second-order equation is

$$\begin{aligned}\frac{dc_1}{dT} =& F\frac{\Omega(c_0)}{E_0{}'(c_0)}\left[2HJ'(c_0) \pm 2bI'(c_0) - \frac{d[cJ(c)J'(c)\Omega(c)]}{dc}\bigg|_{c=c_0} - c_0\left[J'(c_0)\right]^2\Omega(c_0)\right] \\ &+F\Omega'(c_0)c_1,\end{aligned} \tag{31}$$

where $\Omega(c) = \frac{2J(c)c\tanh\frac{H}{W(c)}}{E_0{}'(c)}$, $I(c) = \int_{-H}^{H} \sin\left[A(c)\,\mathrm{sech}\frac{\zeta}{W(c)}\right] d\zeta$ $(\zeta = X - X_0(T))$. Based on the analysis that the first-order equations are identical, the velocities of the positive and negative pulses are taken as

$$\begin{aligned} c^+(T) &= c_0^+(T) + Fc_1^+(T), \\ c^-(T) &= c_0^+(T) + Fc_1^-(T). \end{aligned} \tag{32}$$

Then,

$$\frac{d\left[c_1^+(T) - c_1^-(T)\right]}{dT} = F\Omega'(c_0)\left[c_1^+(T) - c_1^-(T)\right] + 4Fb\frac{I'(c_0)}{E_0{}'(c_0)}\Omega(c_0). \tag{33}$$

From $I'(c) < 0$ and the stability condition $E_0{}'(c_0) < 0$ in Section V B, it follows that

$$c_1^+(T) - c_1^-(T) = 4bF\Omega(c_0)\int_0^T \frac{I'(c_0)}{E_0{}'(c_0)} dT < 0. \tag{34}$$

Combining this with the relation for $A(c)$ gives

$$A^2\left(c^+\right) - A^2(c^-) = -\frac{6K_\theta\left(c^+ + c^-\right)}{(c^+c^-)^2} F\left(c_1^+(T) - c_1^-(T)\right) > 0. \tag{35}$$

Eq. (35) explains why the amplitude of the positive pulse is larger than that of the negative pulse. The corresponding theoretical results are also plotted in Fig. 4(b).

## VI. CONCLUSION

In conclusion, we have found a pathway in vector solitons for a constant force to amplify pulse solitons. In the structure with $b = 0$, the dominant mechanism of amplitude amplification is that the energy of a vector soliton forms a whole, and when the energy of the displacement component increases, the energy of the rotational component must also increase accordingly. This is how energy transfer is achieved through polarization coupling. Furthermore, in the eccentrically loaded structure with $b \neq 0$, polarization coupling provides a nonzero factor for the symmetry-breaking term, thereby enabling asymmetric amplification of positive and negative pulses. Our study demonstrates that polarization coupling, the most fundamental feature of vector deformation, contains a control pathway that distinguishes vector solitons from pure pulse solitons.

## ACKNOWLEDGMENTS

This work was supported by the National Natural Science Foundation of China (Nos. 12272154 and 12532010).

## Appendix A: NUMERICAL METHODS

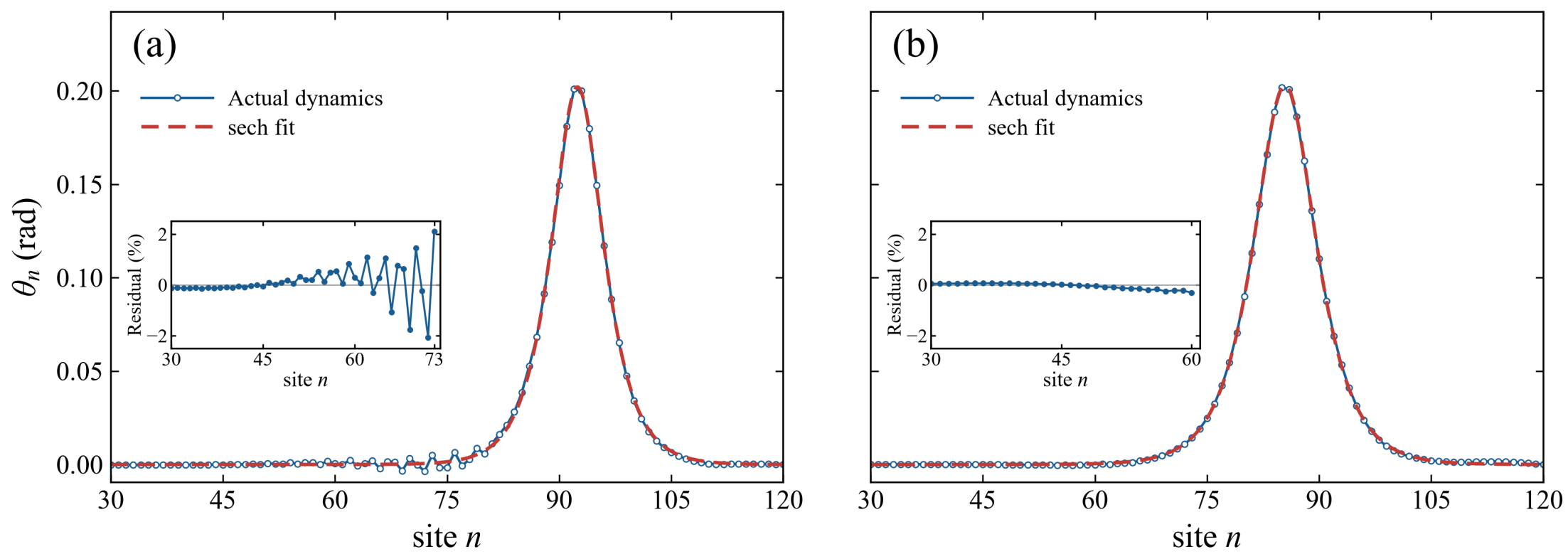

FIG. 6. (a) When $\theta_{\mu_0}$ and $U_{\mu_0}$ are used directly as excitation functions, pronounced phonon radiation appears in the rotational component. (b) $\theta_{\mu_2}(X_{\text{in}}, T)$ and $U_{\mu_2}(X_{\text{in}}, T)$ are used as excitation functions. Phonon radiation is eliminated.

This appendix introduces the numerical method used in this paper, mainly the setting of the soliton excitation functions at the 0-th node. The case of pure force loading at the mass center in Section III is used as an example, and the equations we use to generate the excitation functions are

$$U_{TT} = U_{XX} + \frac{1}{2}\left(\theta^2\right)_X + \mu_j(T)F(X,T), \tag{A1}$$

$$\frac{\theta_{TT}}{\alpha^2} = (K_s - K_\theta)\theta_{XX} - 6K_\theta\theta - 4U_X\theta - 2\theta^3. \tag{A2}$$

$j = 0,1,2$, corresponding to three $\mu_j(T)$ functions with different roles, with $\mu_0(T) = 0$.

$$\mu_1(T) = \begin{cases} 0, & T \le T_{\text{entry}} \\ \frac{1}{2}\left[1 - \cos\left(\pi\frac{T - T_{\text{entry}}}{R}\right)\right], & T_{\text{entry}} < T < T_{\text{entry}} + R \\ 1, & T \ge T_{\text{entry}} + R \end{cases}. \tag{A3}$$

$\mu_2(T) = 1$. The parameters in the equations will be introduced when they are used below. First, the reason why we reconstruct the excitation functions is explained. For $\mu_j(T) = \mu_0(T)$, the analytical solutions for this case have been used as excitation functions to generate solitons with stable profiles [17]. However, when the analytical solutions corresponding to $\mu_0$, namely $\theta_{\mu_0}$ and $U_{\mu_0}$, are applied directly to Eqs. (1) and (2), the mismatch between these unperturbed solutions and the perturbed equations causes phonon radiation. This is shown in the pulse soliton component in Fig. 6(a). To find excitation functions that match the perturbed equations to excite solitons, we use the following three steps.

(1) Set $\mu_j(T) = \mu_1(T)$. The input functions at the 0-th node are the analytical solutions $\theta_{\mu_0}$ and $U_{\mu_0}$. $T_{\text{entry}}$ is set to give the soliton a period of time to stabilize its profile, with $T_{\text{entry}} = 130$. $R$ is set to allow the perturbation term to be applied gradually over a long interval, and $R$ is set to $R = 1000$. Then the time at which the perturbation is fully applied is $T_{\text{full}} = T_{\text{entry}} + R = 1130$. At $T_{\text{full}}$, extract the numerical solutions $\theta_{\mu_1}(X, T_{\text{full}})$, $U_{\mu_1}(X, T_{\text{full}})$ and the corresponding time-derivative data.

(2) Use $\theta_{\mu_1}(X, T_{\text{full}})$, $U_{\mu_1}(X, T_{\text{full}})$, $\dot{\theta}_{\mu_1}(X, T_{\text{full}})$, and $\dot{U}_{\mu_1}(X, T_{\text{full}})$ as the initial conditions for the equations with $\mu_j(T) = \mu_2(T)$, and compute the numerical solutions for $T<0$ to obtain the backward numerical solutions $\theta_{\mu_2}(X,T)$ and $U_{\mu_2}(X,T)$. Since the soliton amplitude changes continuously during propagation, the maximum amplitude at position $X$, $\max\left[\theta_{\mu_2}(X,T)\right]$, varies with $X$. For a target input amplitude of $A_{\text{in}}$, vary $X$ to find where $\max\left[\theta_{\mu_2}(X,T)\right]$ is closest to $A_{\text{in}}$, and denote this $X$ by $X_{\text{in}}$. Store the time-dependent solutions at this reference point, $\theta_{\mu_2}(X_{\text{in}},T)$ and $U_{\mu_2}(X_{\text{in}},T)$, together with the corresponding time-derivative data.

(3) Use $\theta_{\mu_2}(X_{\text{in}},T)$ and $U_{\mu_2}(X_{\text{in}},T)$ as the excitation functions at the 0-th node and apply them to the 0-th node in Eqs. (1) and (2). The actual dynamics generated by the excitation are shown in Fig. 6(b), and the compatible excitation functions prevent phonon radiation.

## Appendix B: ENERGY OF THE SOLITON

The energies of positive and negative pulses are the same, and only the derivation of the energy of a positive pulse is presented here. From Eqs. (9) and (10), the following expressions are obtained

$$\begin{bmatrix} U_T \\ \theta_T \\ U_X \\ \theta_X \end{bmatrix} = \begin{bmatrix} \frac{Jc}{2W}\operatorname{sech}^2 z + \dot{c}\left[\frac{J'}{2}(1-\tanh z) + \frac{JW'}{2W} z \operatorname{sech}^2 z\right] \\ \frac{Ac}{W}\operatorname{sech} z \tanh z + \dot{c}\left[A'\operatorname{sech} z + \frac{AW'}{W} z \operatorname{sech} z \tanh z\right] \\ -\frac{J}{2W}\operatorname{sech}^2 z \\ -\frac{A}{W}\operatorname{sech} z \tanh z \end{bmatrix}, \qquad \text{(B1)}$$

where $z = \frac{X - X_0(T)}{W}$. Substituting Eqs. (9), (10), and (B1) into Eq. (7) gives

$$
\begin{aligned}
E(T) =& \frac{J^2c^2}{4W^2}\int_{-\infty}^{+\infty}\mathrm{sech}^4 z dX + \frac{A^2c^2}{4\alpha^2W^2}\int_{-\infty}^{+\infty}\mathrm{sech}^2 z \tanh^2 z\, dX \\
&+ \frac{3K_\theta}{2}\int_{-\infty}^{+\infty}A^2\mathrm{sech}^2 z dX + \frac{A^2(K_s - K_\theta)}{4W^2}\int_{-\infty}^{+\infty}\mathrm{sech}^2 z \tanh^2 z\, dX \\
&+ \frac{1}{4}\left(A^2 - \frac{J}{W}\right)^2\int_{-\infty}^{+\infty}\mathrm{sech}^4 z dX \\
&+ \int_{-\infty}^{+\infty}\frac{Jc\dot{c}}{W}\,\mathrm{sech}^2 z\left[\frac{J'}{2}(1-\tanh z) + \frac{JW'}{2W}z\,\mathrm{sech}^2 z\right]dX \\
&+ \int_{-\infty}^{+\infty}\frac{Ac\dot{c}}{2\alpha^2W}\mathrm{sech}\, z\tanh z\left[A'\,\mathrm{sech}\, z + \frac{AW'}{W}z\,\mathrm{sech}\, z\tanh z\right]dX \\
&+ \int_{-\infty}^{+\infty}\frac{\dot{c}^2}{4\alpha^2}\left[A'\,\mathrm{sech}\, z + \frac{AW'}{W}z\,\mathrm{sech}\, z\tanh z\right]^2 dX \\
&+ \int_{-\infty}^{+\infty}\dot{c}^2\left[\frac{J'}{2}(1-\tanh z) + \frac{JW'}{2W}z\,\mathrm{sech}^2 z\right]^2 dX,
\end{aligned}
\tag{B2}
$$

Evaluating Eq. (B2), except for the last term, gives

$$
\begin{aligned}
E(T) =& \frac{2K_\theta}{W(c)}\left[\frac{3(K_s-K_\theta)}{c^2} - (K_s - K_\theta) - \frac{2}{\alpha^2}\right] + cJJ'\dot{c} + \dot{c}^2\big(y_1(c)\big) \\
&+ \int_{-\infty}^{+\infty}\dot{c}^2\left[\frac{J'}{2}(1-\tanh z) + \frac{JW'}{2W}z\,\mathrm{sech}^2 z\right]^2 dX,
\end{aligned}
\tag{B3}
$$

where the first term is the instantaneous energy of the soliton, which is defined as

$$
E_0(c) = \frac{2K_\theta}{W}\left[\frac{3(K_s-K_\theta)}{c^2} - (K_s-K_\theta) - \frac{2}{\alpha^2}\right]. \tag{B4}
$$

$y_1(c)$ is given by

$$
y_1(c) = \frac{1}{4\alpha^2}\left(2W(A')^2 + 2AA'W' + \frac{\pi^2+12}{18}\frac{A^2}{W}(W')^2\right), \tag{B5}
$$

where $A = A(c)$, $J = J(c)$, and $W = W(c)$. Only the last term remains in the integration. It is finite in an actual finite structure, but the integral diverges over an infinite domain. This term cannot be treated over an infinite domain. First, it is expanded to separate out the divergent term,

$$
\begin{aligned}
&\int_{-\infty}^{+\infty} \dot{c}^2 \left[\frac{J'}{2}(1-\tanh z) + \frac{JW'}{2W} z \,\mathrm{sech}^2 z\right]^2 dX \\
&= \dot{c}^2 \left[\left(-\frac{J'JW'}{2} + \frac{J^2W'^2}{36W}\left(\pi^2-6\right)\right) + \int_{-\infty}^{+\infty} \left[\frac{J'}{2}(1-\tanh z)\right]^2 dX\right] \\
&= \dot{c}^2 \Bigg[-\frac{J'JW'}{2} + \frac{J^2W'^2}{36W}\left(\pi^2-6\right) + \int_{-\infty}^{+\infty} \frac{J'^2}{4}(1-\tanh z)^2 - \frac{J'^2}{2}(1-\tanh z) dX \\
&\qquad + \int_{-\infty}^{+\infty} \frac{J'^2}{2}(1-\tanh z)\, dX\Bigg] \\
&= \dot{c}^2 \left[-\frac{J'JW'}{2} + \frac{J^2W'^2}{36W}\left(\pi^2-6\right) - \frac{WJ'^2}{2} + \int_{-\infty}^{+\infty} \frac{J'^2}{2}(1-\tanh z)\, dX\right].
\end{aligned}
\tag{B6}
$$

The last term is evaluated in a finite chain as

$$
\int_0^N \frac{J'^2}{2}(1-\tanh z)\, dX = J'^2 \left[\frac{N}{2} - \frac{W}{2} \ln \frac{\cosh\left(\frac{N-X_0}{W}\right)}{\cosh\left(\frac{-X_0}{W}\right)}\right]. \tag{B7}
$$

All our dynamical simulations approximately satisfy the condition that the soliton center is far from both ends, namely $\frac{N-X_0}{W} \gg 1$ and $\frac{X_0}{W} \gg 1$. For $|v| \gg 1$, the relation $\ln\cosh v = |v| - \ln 2 + O\left(e^{-2|v|}\right)$ holds, and Eq. (B7) is approximated as

$$
\int_0^N \frac{J'^2}{2}(1-\tanh z)\, dX \approx J'^2 X_0. \tag{B8}
$$

Therefore, the full expression for the energy can be written as

$$
E(T) = E_0(c) + cJJ'\dot{c} + \dot{c}^2\left(y(c)\right) + \dot{c}^2 J'^2 X_0, \tag{B9}
$$

where $y(c) = y_1(c) - \frac{J'JW'}{2} + \frac{J^2W'^2}{36W}\left(\pi^2-6\right) - \frac{WJ'^2}{2}$.

## Appendix C: DRIVING PULSE SOLITONS WITH A CONSTANT TORQUE

This appendix presents the case of a pure torque acting on the cross units. A torque $rf$ is applied in the soliton core region. $r$ is the arm length of the cross unit minus the magnet radius. This length is chosen to connect the content of Sections IV and V. Therefore, the dynamical equations simulated in this appendix are

$$
\ddot{U}_i = U_{i+1} - 2U_i + U_{i-1} - \frac{1}{2}\left(\cos\theta_{i+1} - \cos\theta_{i-1}\right), \tag{C1}
$$

$$\begin{aligned}\frac{\ddot{\theta}_i}{\alpha^2} =& -K_\theta\left(\theta_{i+1}+4\theta_i+\theta_{i-1}\right)+K_s\cos\theta_i\left(\sin\theta_{i-1}-2\sin\theta_i+\sin\theta_{i+1}\right) \\ & -\sin\theta_i\left[4-\cos\theta_{i-1}-2\cos\theta_i-\cos\theta_{i+1}+2\left(U_{i+1}-U_{i-1}\right)\right]+4bF_i.\end{aligned} \tag{C2}$$

$b=\frac{r}{a}$. Excitations with amplitudes of $A_{\text{in}}=0.2$, $A_{\text{in}}=0.3$, and $A_{\text{in}}=0.4$ were tested. The results for these three excitations are shown in Fig. 7. The results in the figure agree with the predictions of our theory. Constant driving cannot control pulse solitons, and their amplitudes remain nearly unchanged over time.

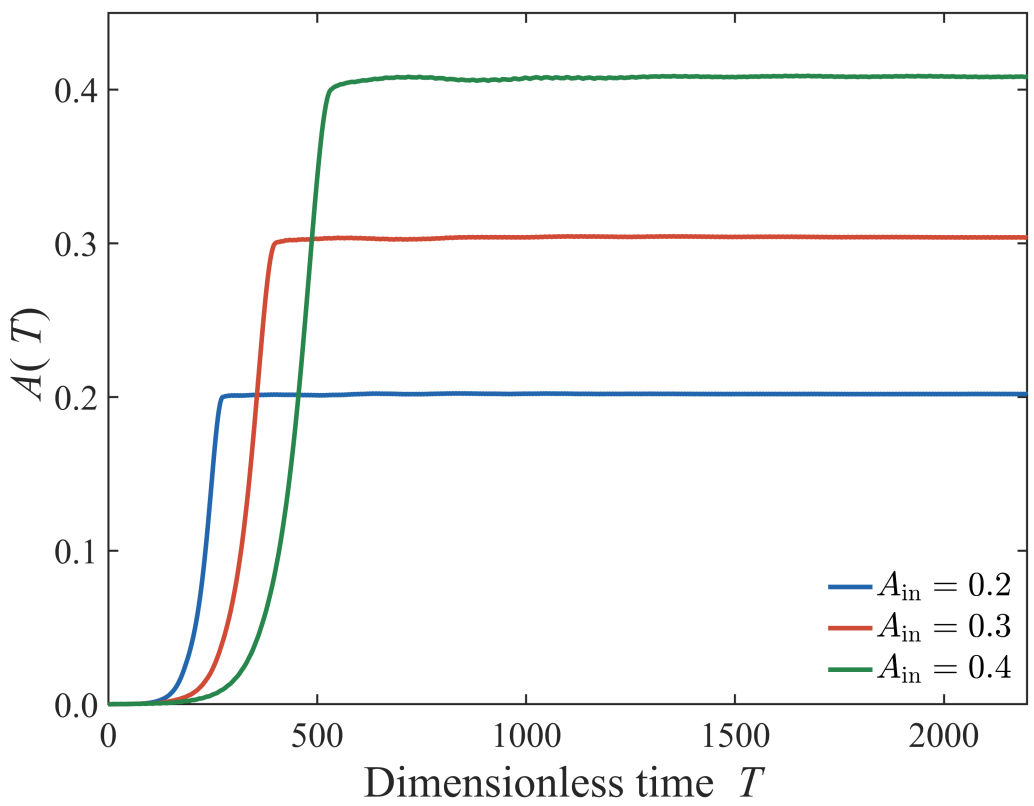

FIG. 7. Plot of the amplitude evolution of a pulse soliton under a constant torque. After the soliton is fully formed, $A(T)$ remains nearly unchanged.

## Appendix D: QUADRATIC FORM OF THE COMOVING HAMILTONIAN

Completing the square in $H_m(c)$ gives

$$\begin{aligned}H_m(c)=\int_{-\infty}^{+\infty}\Bigg[&(U_T+cU_X)^2+\frac{(\theta_T+c\theta_X)^2}{4\alpha^2}+\left(1-c^2\right)\left(U_X+\frac{\theta^2}{2\left(1-c^2\right)}\right)^2 \\ &+\frac{K_s-K_\theta-c^2/\alpha^2}{4}\theta_X^2+\frac{3K_\theta}{2}\theta^2-\frac{c^2}{4\left(1-c^2\right)}\theta^4\Bigg]\,dX.\end{aligned} \tag{D1}$$

Substituting $\theta=\theta_0+\epsilon h_\theta$, $U=U_0+\epsilon h_U$ gives

$$H_m-H_m(\theta_0,U_0)=\epsilon H_m^{(1)}+\epsilon^2 H_m^{(2)}+O\left(\epsilon^3\right), \tag{D2}$$

where $H_m^{(1)}=0$ because the soliton solution corresponds to a stationary point of $H_m$. The expression for $H_m^{(2)}$ is

$$H_m^{(2)}[h] = \int_{-\infty}^{+\infty} \left(h_{U,T} + c h_{U,X}\right)^2 + \frac{\left(h_{\theta,T} + c h_{\theta,X}\right)^2}{4\alpha^2} + (1 - c^2)\left(h_{U,X} + \frac{\theta_0 h_\theta}{1 - c^2}\right)^2 dX + \frac{1}{2}\int_{-\infty}^{+\infty} h_\theta L_\theta\left[h_\theta\right] dX, \tag{D3}$$

where $L_\theta = -\frac{K_s - K_\theta - c^2/\alpha^2}{2}\partial_X^2 + \left(3K_\theta - \frac{3c^2\theta_0^2}{1-c^2}\right)$ and $1 - c^2 > 0$. Under the invertible transformation

$$\begin{bmatrix}\eta_1\\ \eta_2\\ \eta_3\\ \eta_4\end{bmatrix} = \begin{bmatrix} h_{U,T} + c h_{U,X}\\ h_{\theta,T} + c h_{\theta,X}\\ h_{U,X} + \frac{\theta_0 h_\theta}{1 - c^2}\\ h_\theta\end{bmatrix}, \tag{D4}$$

the numbers of negative and zero eigenvalues of the full second-order operator corresponding to $H_m^{(2)}[h]$ are determined by $L_\theta$. The separation of the positive spectrum from zero can be jointly guaranteed by the following two aspects: the first three terms are positive definite, and the positive spectrum of $L_\theta$ is separated from zero. Therefore, the whole problem is transformed into an analysis of the spectral structure of $L_\theta$. The Schrödinger operator $L_\theta$ has one negative eigenvalue, $L_\theta\left(\operatorname{sech}^2\frac{X-cT}{W}\right) = -9K_\theta \operatorname{sech}^2\frac{X-cT}{W}$ , and one zero eigenvalue, $L_\theta\left(\operatorname{sech}\frac{X-cT}{W}\tanh\frac{X-cT}{W}\right) = 0$. The remaining spectrum of $L_\theta$ is continuous and has the range $\left[3K_\theta, +\infty\right)$. Therefore, the full second-order operator corresponding to $H_m^{(2)}[h]$ satisfies the assumptions of stability theory, and its stability region is given by $H_m''(c) > 0$ [27].

## REFERENCES

[1] B. G.-g. Chen, N. Upadhyaya, and V. Vitelli, Nonlinear conduction via solitons in a topological mechanical insulator, Proc. Natl. Acad. Sci. USA **111**, 13004 (2014).

[2] B. Deng, J. R. Raney, V. Tournat, and K. Bertoldi, Elastic vector solitons in soft architected materials, Phys. Rev. Lett. **118**, 204102 (2017).

[3] B. Deng, C. Mo, V. Tournat, K. Bertoldi, and J. R. Raney, Focusing and mode separation of elastic vector solitons in a 2D soft mechanical metamaterial, Phys. Rev. Lett. **123**, 024101 (2019).

[4] L. Jin, R. Khajehtourian, J. Mueller, A. Rafsanjani, V. Tournat, K. Bertoldi, and D. M. Kochmann, Guided transition waves in multistable mechanical metamaterials, Proc. Natl. Acad. Sci. USA **117**, 2319 (2020).

[5] H. Yasuda, L. M. Korpas, and J. R. Raney, Transition waves and formation of domain walls in multistable mechanical metamaterials, Phys. Rev. Appl. **13**, 054067 (2020).

[6] D. Bazeia, L. Losano, and G. J. Olmo, Novel connection between lump-like structures and quantum mechanics, Eur. Phys. J. Plus **133**, 251 (2018).

[7] A. V. Gorbach, S. Denisov, and S. Flach, Energy flow of moving dissipative topological solitons, Chaos **16**, 023125 (2006).

[8] Z. Tang, Q. Xia, H. Chen, S. Fu, and Y. Gao, Bifurcation of the quasi-stationary velocity of strongly discrete transition waves driven by gravity, arXiv:2605.18926.

[9] O. M. Braun, B. Hu, and A. Zeltser, Driven kink in the Frenkel-Kontorova model, Phys. Rev. E **62**, 4235 (2000).

[10] J. R. Raney, N. Nadkarni, C. Daraio, D. M. Kochmann, J. A. Lewis, and K. Bertoldi, Stable propagation of mechanical signals in soft media using stored elastic energy, Proc. Natl. Acad. Sci. USA **113**, 9722 (2016).

[11] N. Nadkarni, A. F. Arrieta, C. Chong, D. M. Kochmann, and C. Daraio, Unidirectional transition waves in bistable lattices, Phys. Rev. Lett. **116**, 244501 (2016).

[12] N. Nadkarni, C. Daraio, R. Abeyaratne, and D. M. Kochmann, Universal energy transport law for dissipative and diffusive phase transitions, Phys. Rev. B **93**, 104109 (2016).

[13] N. R. Quintero, A. Sánchez, and F. G. Mertens, Resonances in the dynamics of $\varphi^4$ kinks perturbed by ac forces, Phys. Rev. E **62**, 5695 (2000).

[14] M. Salerno and Y. Zolotaryuk, Soliton ratchetlike dynamics by ac forces with harmonic mixing, Phys. Rev. E **65**, 056603 (2002).

[15] R. A. Van Gorder, Adiabatic soliton management: Controlling solitary wave motion while keeping the wave envelope unchanged, Phys. Lett. A **446**, 128284 (2022).

[16] R. W. S. Westdorp and H. J. Hupkes, Soliton amplification in the Korteweg-de Vries equation by multiplicative forcing, Commun. Pure Appl. Anal. **24**, 1048 (2025).

[17] B. Deng, P. Wang, Q. He, V. Tournat, and K. Bertoldi, Metamaterials with amplitude gaps for elastic solitons, Nat. Commun. **9**, 3410 (2018).

[18] B. Deng, V. Tournat, P. Wang, and K. Bertoldi, Anomalous collisions of elastic vector solitons in mechanical metamaterials, Phys. Rev. Lett. **122**, 044101 (2019).

[19] J. Veenstra, O. Gamayun, X. Guo, A. Sarvi, C. Ventura Meinersen, and C. Coulais,

Non-reciprocal topological solitons in active metamaterials, Nature **627**, 528 (2024).

[20] J. Li, Y. Yuan, J. Wang, R. Bao, and W. Chen, Propagation of nonlinear waves in graded flexible metamaterials, Int. J. Impact Eng. **156**, 103924 (2021).

[21] B. Deng, J. Li, V. Tournat, P. K. Purohit, and K. Bertoldi, Dynamics of mechanical metamaterials: A framework to connect phonons, nonlinear periodic waves and solitons, J. Mech. Phys. Solids **147**, 104233 (2021).

[22] A. Demiquel, V. Achilleos, G. Theocharis, and V. Tournat, Envelope vector solitons in nonlinear flexible mechanical metamaterials, Wave Motion **131**, 103394 (2024).

[23] A. Demiquel, V. Achilleos, G. Theocharis, and V. Tournat, Gradient catastrophe and Peregrine soliton in nonlinear flexible mechanical metamaterials, Chaos Solitons Fractals **201**, 117201 (2025).

[24] H. Duran, J. Cuevas-Maraver, P. G. Kevrekidis, and A. Vainchtein, Discrete breathers in a mechanical metamaterial, Phys. Rev. E **107**, 014220 (2023).

[25] Q. Zhang, A. V. Cherkasov, C. Xie, N. Arora, and S. Rudykh, Nonlinear elastic vector solitons in hard-magnetic soft mechanical metamaterials, Int. J. Solids Struct. **280**, 112396 (2023).

[26] M. Grillakis, J. Shatah, and W. Strauss, Stability theory of solitary waves in the presence of symmetry, I, J. Funct. Anal. **74**, 160 (1987).

[27] M. Maeda, Stability of bound states of Hamiltonian PDEs in the degenerate cases, J. Funct. Anal. **263**, 511 (2012).